\documentclass[letter]{aa}
\usepackage{amsmath}
\usepackage{physics}
\usepackage{bbold}
\usepackage[normalem]{ulem}
\usepackage{xcolor}
\usepackage{multirow}
\usepackage{tikz}
\usepackage{float}
\usepackage{siunitx}
\usetikzlibrary{shapes.geometric, arrows.meta, positioning}
\usepackage{amsmath,amstext,mathrsfs}
\usepackage{orcidlink}
\usepackage{gensymb}
\usepackage{comment}
\usepackage{siunitx}

\def\apjl{{ApJ}}%
\def\apjs{{ApJS}}%
\def\aap{{A\&A}}%
\newcommand{\mgii}{Mg\,\textsc{ii}}
\newcommand{\rad}{rad\,m$^{-2}$}
\begin{document} 
\title{Magnetised CGM Gas at $z\sim1$ revealed by SPICE-RACS}

   \titlerunning{Magnetic field in CGM}
   \authorrunning{Malik et al,}

   \author{Sunil Malik \orcidlink{0000-0003-4147-626X}\inst{1}\corrauth{smalik@ucm.es  \\ IPARCOS-UCM-26-021}
          \and S.~P.~O'Sullivan\orcidlink{0000-0002-3968-3051}\inst{1}
          \and A.~J.~M.~Thomson \orcidlink{0000-0001-9472-041X}\inst{2}
          \and C.~S.~Anderson \orcidlink{0000-0002-6243-7879} \inst{3}
          \and C.~Van Eck\orcidlink{0000-0002-7641-9946}\inst{3}
          \and L.~Rudnick \orcidlink{0000-0001-5636-7213} \inst{4}
          \and Amit~Seta\orcidlink{0000-0001-9708-0286} \inst{3}
          \and B. M. Gaensler \orcidlink{0000-0002-3382-9558} \and \inst{5,6,7}
          \and Y.~K.~Ma \orcidlink{0000-0003-0742-2006} \inst{8}
          \and Takuya~Akahori \orcidlink{0000-0001-9399-5331}\inst{9}
          \and D.~Alonso-López\orcidlink{0000-0001-9006-0725}\inst{1}
          \and M. Brüggen \orcidlink{0000-0002-3369-7735} \inst{10}
          \and E.~Carretti \orcidlink{0000-0002-3973-8403} \inst{11}
          \and S.~W.~Duchesne \orcidlink{0000-0002-3846-0315} \inst{12}
          \and T.~J.~Galvin \orcidlink{0000-0002-2801-766X} \inst{12}
          \and G.~Heald \orcidlink{0000-0002-2155-6054} \inst{2,12}
          \and O.~Hlinka \inst{13}
          \and A.~Khadir \orcidlink{0009-0001-2196-8251} \inst{6,7}
          \and S. A. Mao \orcidlink{0000-0001-8906-7866} \inst{8}
          \and R.~Omae \orcidlink{0000-0002-7259-1706}\inst{14}
          }

   \institute{Departamento de Física de la Tierra y Astrofísica \& IPARCOS-UCM, Universidad Complutense de Madrid, 28040 Madrid, Spain.
   \and SKA Observatory, SKA-Low Science Operations Centre, 26 Dick Perry Avenue, Kensington WA 6151, Australia.
   \and Research School of Astronomy \& Astrophysics, The Australian National University, Canberra, ACT 2611, Australia.
   \and Minnesota Institute for Astrophysics, University of Minnesota, 116 Church Street SE, Minneapolis, MN 55455, USA.
   \and Department of Astronomy and Astrophysics, University of California, Santa Cruz, 1156 High Street, Santa Cruz, CA 95069, USA.
   \and Dunlap Institute for Astronomy and Astrophysics, University of Toronto, 50 St. George Street, Toronto, M5S 3H4, ON, Canada.
   \and David A. Dunlap Department of Astronomy and Astrophysics, University of Toronto, 50 St. George Street, Toronto, M5S 3H4, ON, Canada.
   \and Max-Planck-Institut f\"ur Radioastronomie, Auf dem H\"ugel 69, 53121 Bonn, Germany.
   \and Mizusawa VLBI Observatory, National Astronomical Observatory of Japan, 2-21-1, Osawa, Mitaka, Tokyo 181-8588, Japan.
   \and Hamburg University, Hamburger Sternwarte, Gojenbergsweg 112, D-21029 Hamburg, Germany.
   \and INAF – Istituto di Radioastronomia, via P. Gobetti 101, 40129 Bologna, Italy.
   \and ATNF, CSIRO Space \& Astronomy, PO Box 1130, Bentley, WA 6102, Australia.
   \and CSIRO Information Management \& Technology, PO Box 883, Kenmore, QLD 4069, Australia.
   \and Department of Social Design Engineering, National Institute of Technology, Kochi College, 200-1 Monobe, Nankoku, Kochi 783-8508, Japan.
  }
   \date{Received 17 March 2026; Accepted 28 April 2026}

\abstract
{
Magnetic fields are expected to permeate the circumgalactic medium (CGM) of galaxies, yet direct constraints at high redshift remain limited by the lack of high-quality Faraday rotation measure (RM) data.
Using the RMs from SPICE-RACS DR2 combined with the DESI DR1 quasar catalogue, we compile the largest sample to date of 2483 quasar sightlines with associated RMs, including 612 with intervening \mgii\ absorbers tracing foreground galaxies and 1871 control sightlines without \mgii\ absorbers.
After subtracting the Galactic RM contribution and restricting the analysis to sightlines with low Milky Way H\textsc{i} column density and H$\alpha$ intensity, we obtain a foreground-cleaned sample of 757 quasars (191 \mgii\ / 566 control) spanning redshifts $0.13<z<3.45$.
In this foreground-cleaned sample, \mgii\  sightlines exhibit a $4.5\sigma$ excess in the residual RM dispersion of $4.13 \pm 0.91$~\rad~relative to the control sample, at a median absorber redshift of $z \sim 1.14$. 
This implies model-dependent CGM magnetic field strengths of $\sim0.4 - 0.8\, \mu$G over projected radii of 20 -- 150 kpc.
This indicates that substantial CGM magnetisation was already established by $z\sim1$, enabling new constraints on the growth and amplification of magnetic fields in galaxy halos over cosmic time.
}
\maketitle
\nolinenumbers
\section{Introduction}
\label{introduction}
%\vspace{-0.5em}
The circumgalactic medium (CGM) is a vast, multiphase reservoir of magnetised gas extending beyond galactic disks~\citep{Tumlinson_2017}.
%, and serves as the primary interface between galaxies and the intergalactic medium %and regulating gas accretion
Magnetic fields in and around galaxies can influence CGM dynamics by regulating gas cooling, cosmic-ray confinement, 
and angular-momentum transport. Observations of polarised synchrotron emission and Faraday rotation reveal that coherent and turbulent magnetic fields extend beyond galactic disks, plausibly tracing outflows, inflows, and feedback-driven winds~\citep{kronberg_absorption_1982,kronberg_global_2008,Bernet2008,Farnes_2014, prochaska_zheng_2019, ravi2019, Shah2021MNRAS, Kovacs2024A&A, Khrykin2025}. Characterising the strength, structure, and origin of magnetic fields in the CGM of high-redshift ($z$) galaxies is therefore essential for a complete understanding of galaxy evolution and cosmic magnetism. 

While magnetic fields with strengths of a few $\mu$G are well established in galaxy disks and inner halos, their properties in the extended CGM ($\sim$20–150~kpc) remain poorly constrained. Edge-on galaxy surveys such as Continuum Halos in Nearby Galaxies, an EVLA Survey (CHANG-ES), reveal ordered, X-shaped halo fields and regular magnetic components within $\sim$5–15~kpc of galactic disks, providing a qualitative framework for magnetised halos extending into the CGM \citep{Wiegert2015}. More direct probes are provided by Faraday rotation measures (RMs) from background radio sources, either as a function of impact parameter or through absorber-selected samples. Early studies of Mg \textsc{ii} systems \citep{Bernet2008,Bernet_2010,bernet_extent_2013,Farnes_2014,kim_2016} reported substantial RM excesses of $\sim10$~\rad\ without accounting for Milky Way contamination, while subsequent investigations \citep{malik_role_2020,Lan_and_Prochaska_2020,Burman_2024}, based on NRAO VLA Sky Survey (NVSS) RMs with large uncertainties and poorly constrained Galactic RM (GRM) modelling, inferred extragalactic RM signals implying $\gtrsim\mu$G magnetic fields at tens of kiloparsecs. In contrast, modern RM grids, including recent Low Frequency Array (LOFAR) based analyses, find excesses of $\lesssim4.0$~\rad\, implying $\lesssim \mu$G field strengths within $\sim$50–100~kpc in low-$z$ galaxies \citep{Heesen_CGM_magfield}. At $z\sim 0.5$, lensed quasars have also been used to probe magnetic fields in the galaxy disks~\citep{Mao2017NatAs}, and halos~\citep{Bockmann_2023,Kovacs2026A&A}, while at $z\sim2.6$ by~\cite{Geach2023Natur, Roo2025MNRAS} reported a high magnetic field of $\sim 500 \ \mu$G in the molecular disk. Therefore, due to the disparities in these studies, a detailed investigation is needed using large samples of precise RMs. 
%Additionally, Fast Radio Bursts (FRBs) provide complementary constraints by jointly probing dispersion measure and RM through individual halos, indicating significant magnetisation in the CGM~\citep{prochaska_zheng_2019, ravi2019, Kovacs2024A&A, Khrykin2025}. Cosmological Magnetohydrodynamics simulations predict CGM magnetic fields shaped by feedback and accretion, with $\mu$G-level strengths near disks declining to $\lesssim10^{-2}$–$10^{-1}$~$\mu$G at large radii~\citep{ramesh2023MNRAS}. The predicted RM signals lie near current observational limits, motivating targeted absorber-based analyses as a key way to probe the magnetic field in high-$z$ galaxies.

In this letter, we detect a statistically robust RM excess for \mgii-absorber galaxies by compiling a sample of high-precision RMs of quasars, which is three times larger than previous studies, using the Australian SKA Pathfinder (ASKAP) observations, to probe magnetised CGM environments at high redshift ($z\sim1$). To mitigate Milky Way contamination, we employ a bespoke annulus-based GRM estimation method and remove sightlines passing through regions of high gas density and ionisation in the Galactic foreground. This carefully filtered RM sample allows better isolation of the extragalactic signal and constrains the magnetisation of the CGM in \mgii-selected galaxies.
%\vspace{-0.9em}
\section{Data}
\label{sec:data}
%\vspace{-0.5em}
For the RM measurements, we utilize the recently developed Spectra and Polarisation in Cutouts of Extragalactic Sources Rapid ASKAP Continuum Survey (SPICE-RACS\footnote{SPICE–RACS is a joint initiative of the POSSUM Collaboration (possum-survey.org) and the ASKAP Observatory.}) Data Release 2, which covers the entire southern sky and extends up to a declination of $+49^\circ$ in the north \citep{Alec2026PASA}. The survey operates over a frequency range of 800–1088 MHz with an average angular resolution of $\sim 15^{\prime\prime}$. With an rms noise of $\sim 200 \ \mu\mathrm{Jy}$, approximately 5 million radio components are reported. By applying cuts of \texttt{snr\_polint > 8}, \texttt{fracpol > 0.005}, and a flag as defined in DR2 based on poor convergence in fitting the Stokes I spectrum, \texttt{stokesI\_fit\_flag\footnote{\url{https://github.com/CIRADA-Tools/RM-Tools/wiki/RMsynth1D\#stokes-i-modelling}} 
%\footnote{As defined in the SPICE-RACS DR2 catalog.} 
< 5}  and considering single RM components within $10^{\prime\prime}$, a catalogue of $\sim 250{,}000$ unique RM components has been produced. This is the RM sample used in our subsequent analysis.

%a flag as defined in DR2 based on poor convergence in fitting the Stokes I spectrum, stokesI (as defined in the SPICE-RACS DR2 catalog)

To find the RMs associated with quasars, we cross-match with the Dark Energy Spectroscopic Instrument (DESI) DR1 spectroscopic quasar catalogue~\citep{desi_dr1}. This catalogue contains $\sim 1.6$ million sources up to a redshift of $z \sim 5.0$, with a spectral resolution ranging from $R \sim 2000$–$5000$ over the wavelength range $3000$–$9800 \ \si{\angstrom}$, and provides sub-arcsecond spatial resolution, with a sky coverage of $-18\degree < \delta < +84\degree$. 
The completeness within this coverage is not uniform (see Fig.~\ref{hi_clean_sample}). 
%(see \cite{desi_dr1}, fig.~3, top panel).

We cross-matched the SPICE-RACS and DESI catalogue positions using a matching radius of $3^{\prime\prime}$, considering the angular resolutions of DESI and SPICE-RACS. This results in a sample of 2483 unique sources that are well characterised at both optical and radio wavelengths, and with Galactic latitude $|b| > 20\degree$ (to minimise the galactic plane contribution). 
%The sample has the following properties: (i) redshift range $0.13 < z < 4.6$, (ii)  $-18^\circ <$ Dec.~$< +49^\circ$, (iii) RM values spanning $-200$ to $+150$ \rad, and (iv) polarisation fractions in the range $0.3$–$20\%$.
 
To classify the sightlines based on the presence of foreground galaxies, we use the \mgii \ doublet absorption lines at wavelengths $\lambda\lambda  2796 \ \& \ 2803 \ \si{\angstrom}$ detected in the quasar spectra~\citep{Napolitano2023AJ}. The presence of these absorbers is widely regarded as a reliable proxy for intervening foreground galaxies~\citep{Kacprzak_2008,Chen_2010}. Some of the sightlines have more than one absorber along the line of sight. The RMs of the sightlines with and without \mgii \ absorbers are utilised in constraining the magnetised plasma in the CGM of these galaxies.  
%\vspace{-1.5em}
\subsection{Removal of the Milky Way RM contribution}
%\vspace{-0.5em}
To estimate the foreground GRM for each RM component, we adopt an annulus-based method \citep{Anderson2024MNRAS} which is discussed in the Appendix ~\ref{grm_appendix}, and obtain the residual RM defined as RRM = RM $-$ GRM. In Fig.~\ref{sf}, we use RM structure functions to show that the GRM subtraction using the annulus method effectively removes the majority of the scale-dependent GRM from the RM.

%To estimate the foreground GRM for each RM component, we adopt an annulus-based method \citep{Anderson2024MNRAS}. For each target source, we construct an annulus centered on its position with an inner radius of $r_{\mathrm{inner}} = 0.2^\circ$ (at $z=1.0$, this corresponds to $\sim5.9$ Mpc using Planck cosmology~\citep{Planck2020A&A}), to avoid both self-contamination and contamination from the CGM of the foreground galaxies, and then select the 20 nearest RMs from the full sample. %reference sample. To minimise the influence of outliers, we discard RMs exceeding $5\sigma_{RM}$, where $\sigma_{RM}$ is the standard deviation of the selected 20 RMs, and assign their median value as the GRM at that location. The GRM uncertainty is estimated as $\sigma_{\mathrm{GRM}} = \sigma_{\mathrm{MAD}}/\sqrt{N}$, where $\sigma_{\mathrm{MAD}}$ is the median absolute deviation–based standard deviation and $N=20$ (i.e.~the statistical uncertainty). These GRM estimates are used in our analysis to remove the Milky-Way contribution in the observed RM, and obtain residual RM defined as RRM = RM $-$ GRM. We have also compared our GRM estimates with those from  \cite{Hutschenreuter_2022} (see Appendix ~\ref{grm_appendix}). In Appendix~\ref{sf} we show using RM structure functions that the GRM subtraction using the annulus method effectively removes the majority of the scale-dependent GRM from the RM.  
%\vspace{-1.5em}
\subsection{Clean RM sample selection using HI and H$\alpha$ cuts}
%\vspace{-0.5em}
\label{clean_sample}
After correcting the RMs for the Milky Way contribution using the GRM, it is essential to further examine the potential residual contamination arising from the high-density regions of the interstellar medium. We find that the RRM dispersion is strongly correlated with both H\textsc{i} (using the HI4PI map \citep{HI4PI2016}) and H$\alpha$ intensity (from the WHAM map \citep{Haffner2003}) as shown in Fig.~\ref{sigma_mad_hi_halpha_main}, indicating that the GRM is not removed completely. Therefore, the high-density warm ionised and neutral phases of the Milky Way, traced respectively by H$\alpha$ and H\textsc{i} emission, contribute to the excess scatter in the RRM values.
\begin{figure}[h]
    \centering
    \includegraphics[width=0.88\linewidth,clip=true,trim=0.4cm 0cm 0.5cm 0cm]{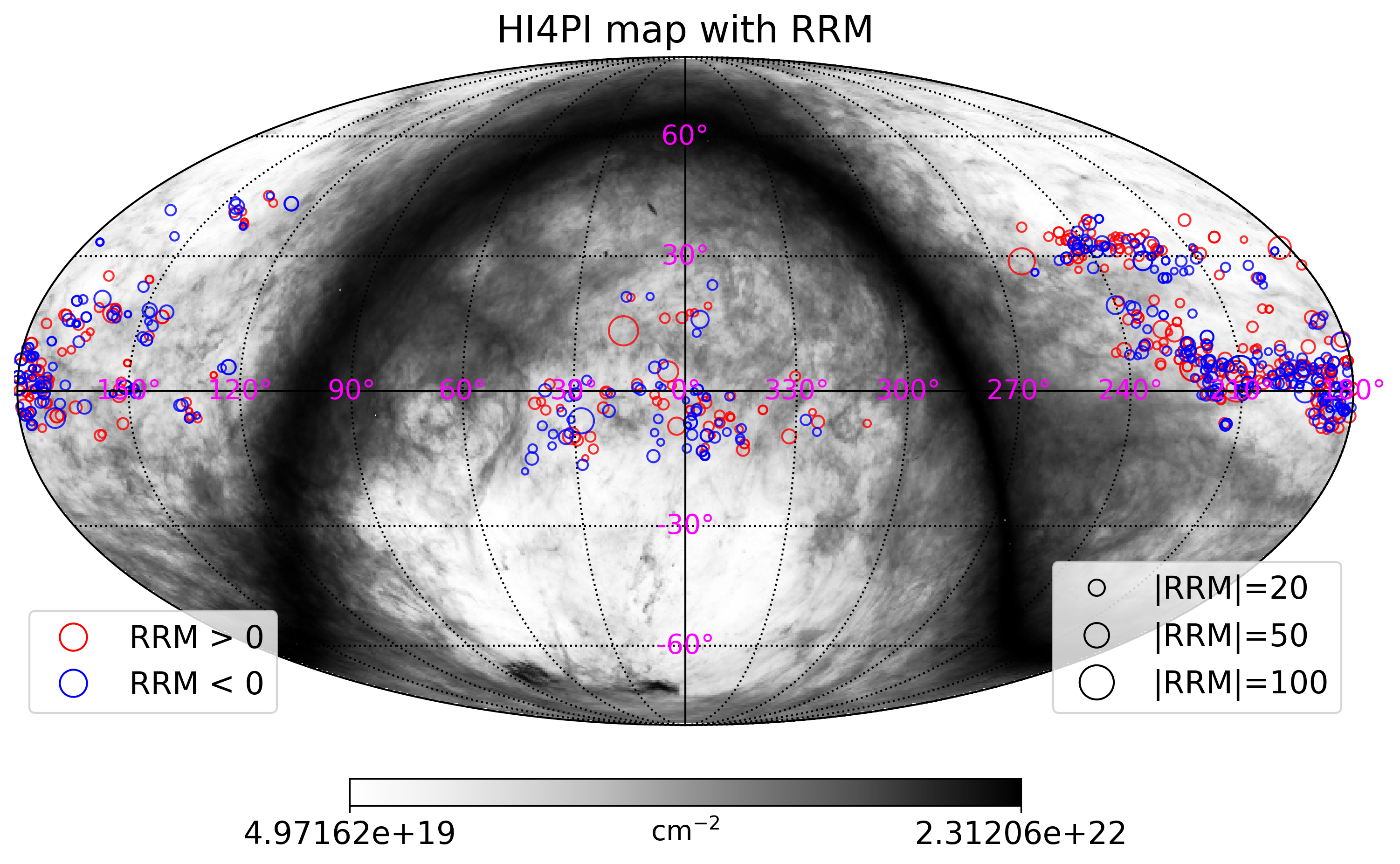}
    \caption{All-sky map of the Galactic neutral hydrogen (H\textsc{i}) column density from the HI4PI survey, overlaid with the clean RRM sample (757 sources). Positive and negative RRMs are shown with red and blue circles, respectively, with the circle size proportional to $|{\rm RRM}|$. Sky coverage of the sources is limited in declination by the SPICE-RACS upper limit of $\sim+49\degree$ and the lower limit  of $-18\degree$ is from DESI. In addition, the source scarcity in some low HI density regions is due to the patchy coverage in the DESI DR1 observations (see \cite{desi_dr1}, fig.~3, top panel). }
    \label{hi_clean_sample}
\end{figure}
%\vspace{-0.5em}
To minimise these effects and obtain a reliable set of sightlines, we have selected sightlines with H\textsc{i} column density and H$\alpha$ intensity below $3.5 \times 10^{20} \ \textrm{cm}^{-2}$ and 1~Rayleigh (R), respectively (for details see Appendix~\ref{hi_alpha_appendix}). Applying these cuts yielded the final clean sample of 757 sources spans over redshifts $0.13<z<3.45$ (as shown in Fig.~\ref{hi_clean_sample}) having 191 and 566 sightlines with and without \mgii \ absorbers, respectively (see Fig.\ref{z_hist} for redshift distribution). This sample represents a minimally contaminated set of sightlines for probing the extragalactic RM signal and for assessing the contribution of intervening Mg\,\textsc{ii} absorbers to the observed RRM dispersion. 

% To minimise these effects and obtain a reliable set of sightlines, we have applied the following filtering criteria; (i) exclusion of sightlines with high H\textsc{i} column density higher than a threshold of $3.5 \times 10^{20} \ \textrm{cm}^{-2}$, (for details see Appendix~\ref{hi_alpha_appendix}) and (ii) selection of sightlines with H$\alpha$ intensity below 1~\textbf{Rayleigh} (R) based on the WHAM survey (for details see Appendix~\ref{hi_alpha_appendix})

\begin{figure}[h]
    \centering
    \includegraphics[width=0.9\linewidth,clip=true,trim=0.2cm 0cm 0.2cm 0.2cm]{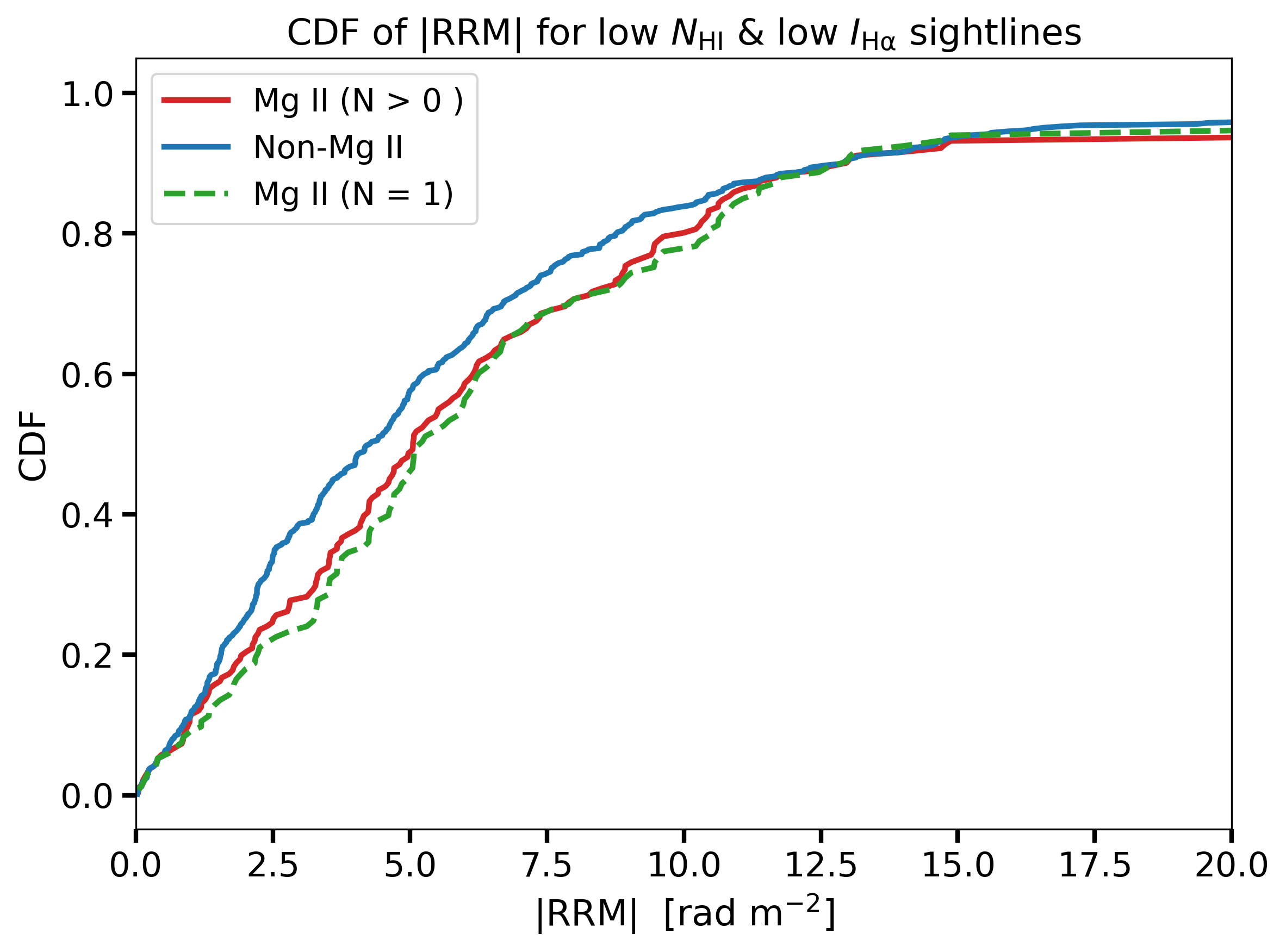}
    \caption{CDF of $|\rm RRM|$ for the subsamples with and without foreground \mgii \ absorbers.}
    \label{cdf_rrm}
\end{figure}
%\vspace{-1.9em}
\section{Results}
\label{sec:method}
%\vspace{-0.5em}
\subsection{RRM Excess in Mg\,\textsc{ii} absorbers}
\label{excess_1}
%\vspace{-0.5em}
To demonstrate the statistical difference between the RRM distributions of the subsamples with and without \mgii \ absorbers, we plot the cumulative distribution function (CDF) of $|\mathrm{RRM}|$ in Fig.~\ref{cdf_rrm}. The CDF clearly shows a systematic shift toward higher RRM values for the Mg \textsc{ii} absorber subsample.
To quantify the RM contribution of the foreground \mgii~absorbers in the redshift range of $0.39<z<2.32$, we computed the excess RRM dispersion associated with the Mg\,\textsc{ii} absorbers as
\begin{equation}
\scriptsize
    \sigma_{\rm excess} = \sqrt{\sigma_{\rm MAD,\,MgII}^2 - \sigma_{\rm MAD,\,noMgII}^2},
    \label{excess}
\end{equation}
where $\sigma_{\rm MAD,\,MgII}$ and $\sigma_{\rm MAD,\,noMgII}$ represent the median absolute deviation of RRM for the subsamples with and without Mg\,\textsc{ii} absorbers, respectively. The uncertainties are propagated to RRM from RM and GRM, and further uncertainties on $\sigma_{\rm MAD}$ of each sample by taking rms of $\sigma_{\rm MAD}$ of simulating $10^3$ realizations of the dataset.
The results are summarised in Table~\ref{tab:rrm_mgii_excess_final}. An excess RRM dispersion of $4.13\pm0.91$ rad~m$^{-2}$ at $4.5\sigma$ was found (uncertainty calculated using equation in footnote)\footnote{\scriptsize
$\delta{\sigma_{\text{excess}}} =
\sigma_{\text{excess}}^{-1} ((\sigma_{\text{MAD, MgII}})^2 (\delta{\sigma_{\text{MAD,MgII}}})^2 \\
 + (\sigma_{\text{MAD,noMgII}})^2 (\delta{\sigma_{\text{MAD,noMgII}}})^2)^{1/2}$}. For the sample with $N=1$, we note that the excess is $5.22 \pm 0.90$~rad m$^{-2}$, which is higher ($\sim 1$ \rad \,) than the value for the sample with $N>0$, which requires further investigation.
This RRM excess is comparable to that found by \citet{Heesen_CGM_magfield} for nearby galaxies using the high-precision LOFAR RMs \citep{osullivanDR2rmgrid}. 
We note that the excess in the total sample before applying the HI and H$\alpha$ cuts is statistically insignificant, with a value of $1.73 \pm 1.65$~\rad\ ($1\sigma$). 

\begin{figure}[h]
    \centering
    \includegraphics[width=0.9\linewidth,clip=true,trim=0.5cm 0.5cm 0.5cm 0.4cm]{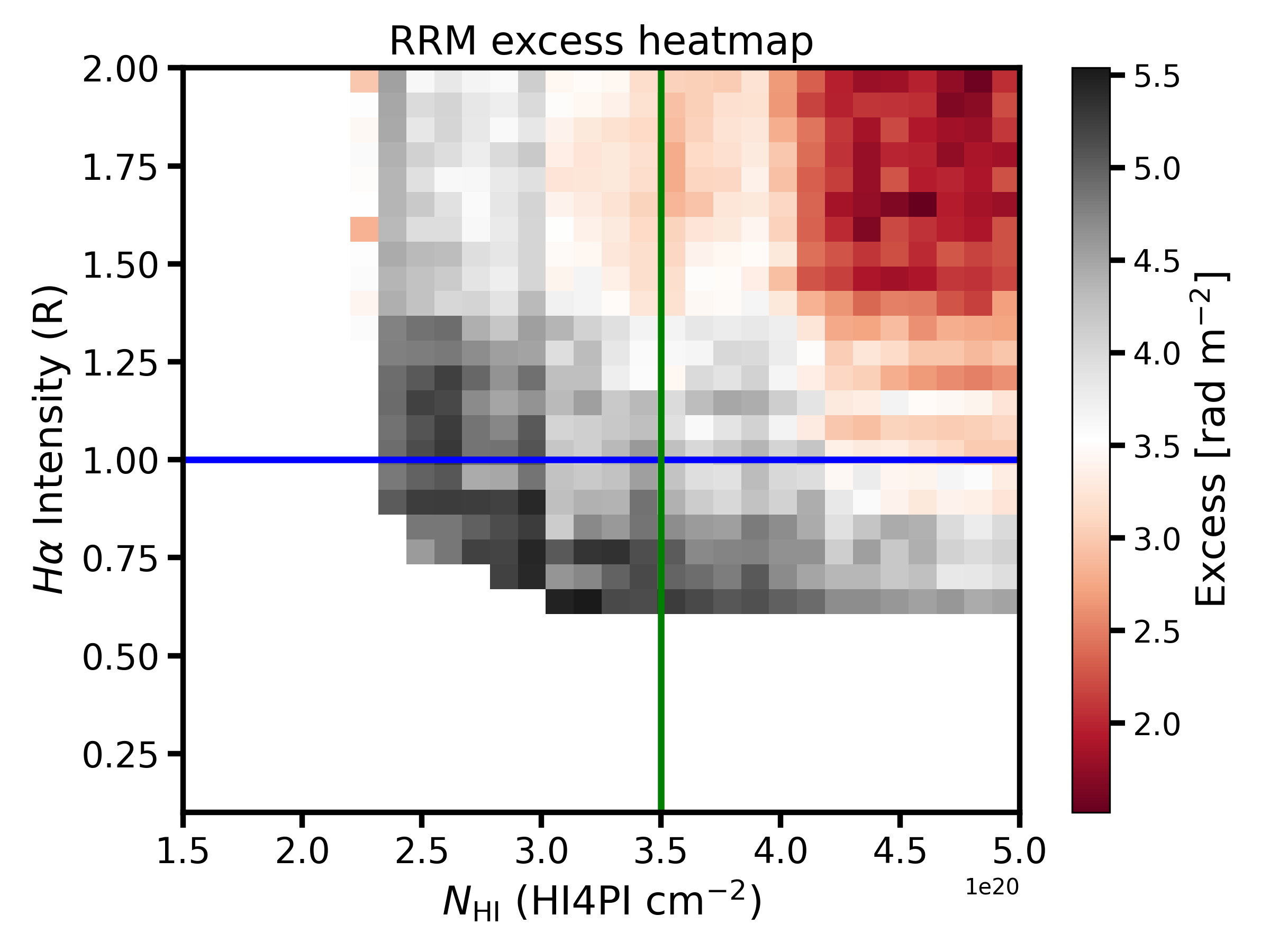}
    \caption{Map of the RRM excess for various cutoff limits of H\textsc{i} column density and H$\alpha$ intensities to illustrate the variation of $\sigma_{\mathrm Excess}$. We retain excess values when both subsamples (with and without Mg II absorbers) contain a minimum of 100 sightlines. The vertical green and horizontal blue lines mark the thresholds H\textsc{i} $= 3.5 \times 10^{20}\ {\rm cm}^{-2}$ and H$\alpha$ = 1.0 R, respectively, adopted to compute the values reported in the table ~\ref{tab:rrm_mgii_excess_final}. The statistical significance heat map is shown in Fig.~\ref{siginficane_excess_hi_halpha_main}.}
    \label{rrm_excess_hi_halpha_main}
\end{figure}

We further tested the significance of the RRM excess by randomly drawing multiple subsamples of 191 sources from the non-absorber sample. The resulting excess has a median value of $\sigma_{\rm excess}$ is $4.18\pm1.06$~\rad\, with a range of $3.56\pm1.3$ to $5.36\pm0.76$~\rad\ . In Fig.~\ref{rrm_excess_hi_halpha_main}, we show that the RRM excess is insensitive to the exact H\textsc{i} and H$\alpha$ cuts, where the excess remains across a wide range of values. This is also true for the significance of the excess (Fig.~\ref{siginficane_excess_hi_halpha_main}).
\vspace{-1.0em}
\begin{table}[h!]
\centering
\scriptsize
\caption{RRM dispersion ($\sigma_{\rm MAD}$) and excess for Mg\,\textsc{ii} subsamples after all selection criteria.}
\label{tab:rrm_mgii_excess_final}
\setlength{\tabcolsep}{4pt}
\begin{tabular}{lccc}
\hline
\hline
\multirow{2}{*}{Mg\,\textsc{ii} Subsample} & $\sigma_{\rm MAD}$ & No of Sightlines & $\sigma_{\rm Excess}$ \\
 & [rad\,m$^{-2}$] & & [rad\,m$^{-2}$] \\
\hline
$N=0$   & $6.48 \pm 0.27$ & 566 & -- \\
$N>0$   & $7.69 \pm 0.43$ & 191 & $4.13 \pm 0.91$ \\
$N=1$   & $8.33 \pm 0.53$ & 133 & $5.22 \pm 0.90$ \\
%$N>2$   & $8.95 \pm 1.39$ & 21  & $6.17 \pm 2.03$ \\
\hline
\end{tabular}
\end{table}
We also performed a Bayesian model comparison of the RRM to test whether sightlines with Mg\,\textsc{ii} absorbers exhibit excess RM relative to the non-Mg\,\textsc{ii}  subsample. Assuming that the RRMs are Gaussian-distributed with zero mean, we model the observed scatter as the quadrature sum of measurement uncertainties and a dispersion, characterised using the Gaussian-equivalent $\sigma_{\rm MAD}$. Two hypotheses are considered: $H_0$, in which the Mg\,\textsc{ii} and non-Mg\,\textsc{ii} sightlines have similar dispersion, and $H_1$, in which the two samples have different dispersions. Adopting Jefferys priors on the dispersion parameters and marginalising over them, we compute the Bayesian evidence for both hypotheses and evaluate the Bayes factor.  We find a Bayes factor of 41.6, providing very strong (decisive) evidence that the sample with Mg \textsc{ii} has an RRM excess. In Appendix~\ref{spectral}, we show that the RRM excess is not significantly affected by splitting the sample by radio spectral index (i.e.~used by previous studies as a proxy for source-compactness).  
\vspace{-0.5em}
%\vspace{-2.8em}
\subsection{Estimate of the magnetic field strength}
%\vspace{-0.5em}
To infer the physical CGM magnetic field, we simultaneously combine (i) energy equipartition between the thermal gas and magnetic field, and (ii) a turbulent Faraday-screen model for the RM dispersion. We assume approximate pressure balance in the CGM, $n_{\rm tot}kT \approx \frac{B^2}{8\pi}$. 
where $n_{\rm tot}$ and $T$ are the total gas density (hydrogen and helium) and temperature of the CGM, respectively. We relate the electron density to the total gas density through an ionization fraction $x_e$ such that $n_e = x_e n_{\rm tot}$. 
Adopting $x_e = 0.8$ and $T = 10^4$~K (as appropriate for predominantly photoionised cool CGM gas, see ~\citep{Dutta2024MNRAS}), we obtain $n_e(B) = x_e\,\frac{B^2}{8\pi kT}$.

We model the Faraday rotation as arising from a turbulent magnetised plasma, for which the RM dispersion is given by~\citep{Gaensler2001ApJ,seta_federrath_2021};
\begin{equation}
\scriptsize
\frac{\sigma_{\rm RM}}{\mathrm{rad\ m^{-2}}} = \frac{812}{2\sqrt{3}(1+z)^2}\,
\left(\frac{n_e}{\mathrm{cm}^{-3}}\right)
\left(\frac{B}{\mu\mathrm{G}}\right)
\sqrt{\left(\frac{L l}{\mathrm{kpc^2}}\right)},
\end{equation}
where $B$ denotes the turbulent magnetic field strength, $L$ is the CGM path length, $l$ is the correlation length, and $z$ is the absorber redshift. Substituting the equipartition relation $n_e(B)$ into the above equation, we solve for the magnetic field as a function of CGM path length, $B(L)$, for a fixed observed excess $\sigma_{\rm RM}$. We adopt $l = L/10$, and marginalise over the absorber redshift range $z=0.37$ to $2.3$ using a Monte Carlo approach with $10^3$ samples.

For a plausible impact parameter of line of sight and CGM extent of $L = 20 - 150$ kpc, we infer magnetic field strength ranges 
$B \approx 0.4 \,{\rm to} \, 0.8 ~\mu$G. This corresponds to equipartition electron densities of $\approx 10^{-2}~\mathrm{cm^{-3}}$ to $10^{-3}~\mathrm{cm^{-3}}$. It is also possible that the CGM has a higher temperature and a relatively low ionisation fraction, which would result in lower $n_{e}$.
%\vspace{-1.5em}
\section{Discussion and Summary}
%\vspace{-0.6em}
\label{sec:discussion}
The excess RRM detection in high-$z$ \mgii \ absorbers provides robust evidence for magnetised gas in the CGM, which possibly results from galactic outflows preferentially aligned with the host galaxy minor axis \citep{Heesen_CGM_magfield}. Several previous studies \citep{Bernet2008,Farnes_2014,kim_2016} have used RM measurements to infer magnetic fields in absorbers; however, their results were severely affected by GRM contamination. Subsequent attempts \citep{Basu2018MNRAS, malik_role_2020,Lan_and_Prochaska_2020, Shah2021MNRAS, Heesen_CGM_magfield,Burman_2024} used RRMs, with GRM subtractions based on relatively simplistic or incomplete foreground modelling. This left several discrepancies related to the large uncertainties of NVSS RMs, limited sample sizes, and incomplete treatment of Galactic foregrounds. In this study, we use RM measurements that are an order of magnitude more precise than NVSS, with a sample size approximately three times larger, and with significantly improved radio–optical positional matching. Moreover, we subtract the GRM using an improved annulus-based method and exclude sightlines intersecting dense neutral and ionised Galactic regions, yielding a substantially more robust RRM sample. We find that the RRM excess in \mgii~absorbers is significant, and implies that normal galaxies host a magnetised CGM which is larger than predicted by simulations \citep[e.g.][]{ramesh2023MNRAS}.

Several caveats remain in translating the observed RRM dispersion into physical magnetic field strengths. Such conversions depend on poorly constrained CGM properties, including electron density, path length, and magnetic-field coherence, which may vary substantially with halo mass, redshift, and environment. Residual Galactic RM structure, despite stringent H\textsc{i} and H$\alpha$ cuts, may still contribute to the scatter, while source-dependent effects such as Faraday depolarisation can modulate the observed RRM independently of the intervening medium. Moreover, Mg \textsc{ii}-selected absorbers preferentially trace cool, metal-enriched gas and may not be representative of the full CGM volume, potentially biasing magnetic field estimates toward denser or more strongly magnetised regions. The estimates in Section~\ref{excess_1} are influenced by these effects, and a more detailed investigation of the radial variation and redshift evolution of the magnetic field in these absorbers will be presented in a follow-up paper. 

Despite these limitations, this analysis using a conservative and homogeneous methodology successfully isolates the extragalactic RM signal. In summary, 
\vspace{-0.5em}
\begin{itemize}
\item  We compiled a sample of 2483 quasars with well-characterised radio and optical properties, extending up to $z \sim 4.0$, of which 757 sightlines (191 with and 566 without \mgii) had low foreground Galactic HI density and H$\alpha$ intensity. We found a decisive (Bayes factor of $\sim42$) excess extragalactic RM associated with the \mgii \ absorbers of $4.13\pm0.91$~\rad \ ($4.5\sigma$).   %\textcolor{blue}{why 4.0 and not 4.5sigma?} 

\item We emphasize that to mitigate Galactic contamination by GRM removal alone is insufficient. Additional cuts of H\textsc{i} $< 3.5 \times 10^{20}\ {\rm cm}^{-2}$ and H$\alpha$ intensity $< 1$~R were needed to minimise the Milky Way RM contamination from high gas density and ionised regions, without which the RRM excess is not statistically significant (<$1\sigma$). 

\item We estimated the CGM magnetic field $B\sim 0.4 -0.8 \, \mu$G strength at typical radii between 20 to 150 kpc, assuming energy equipartition and nominal parameter values that require further detailed investigation. 

\end{itemize}

This study provides a reliable assessment of CGM magnetisation at high redshift and establishes a framework that is readily extendable to forthcoming high-density RM grids from the Polarisation Sky Survey of the Universe's Magnetism (POSSUM)~\citep{Gaensler2025PASA} and the SKA, which will be essential for resolving the coherence scale, geometry, and redshift evolution of magnetic fields in the circumgalactic medium.

\begin{acknowledgements}
We thank the anonymous referee for constructive comments that significantly improved the paper. SM, SPO and DAL acknowledge support from grant PID2023-146372OB-I00, funded by MICIU/AEI/10.13039/501100011033 and by ERDF, EU. SPO acknowledge support from the Comunidad de Madrid Atracción de Talento program via grant 2022-T1/TIC-23797.
C.S.A. acknowledges funding from the Australian Research Council in the form of Australian Future Fellowship FT240100498. AS is supported by the Australian Research Council through the Discovery Early Career Researcher Award (DECRA) Fellowship (project~DE250100003) funded by the Australian Government and the Australia-Germany Joint Research Cooperation Scheme of Universities Australia (UA--DAAD, 2025--2026).

This scientific work uses data obtained from Inyarrimanha Ilgari Bundara / the Murchison Radio-astronomy Observatory. We acknowledge the Wajarri Yamaji People as the Traditional Owners and native title holders of the Observatory site. The Australian SKA Pathfinder is part of the Australia Telescope National Facility (https://ror.org/05qajvd42) which is managed by CSIRO. Operation of ASKAP is funded by the Australian Government with support from the National Collaborative Research Infrastructure Strategy. ASKAP uses the resources of the Pawsey Supercomputing Centre. Establishment of ASKAP, the Murchison Radio-astronomy Observatory and the Pawsey Supercomputing Centre are initiatives of the Australian Government, with support from the Government of Western Australia and the Science and Industry Endowment Fund.

This research used data obtained with the Dark Energy Spectroscopic Instrument (DESI). DESI construction and operations is managed by the Lawrence Berkeley National Laboratory. This material is based upon work supported by the U.S. Department of Energy, Office of Science, Office of High-Energy Physics, under Contract No. DE–AC02–05CH11231, and by the National Energy Research Scientific Computing Center, a DOE Office of Science User Facility under the same contract. Additional support for DESI was provided by the U.S. National Science Foundation (NSF), Division of Astronomical Sciences under Contract No. AST-0950945 to the NSF’s National Optical-Infrared Astronomy Research Laboratory; the Science and Technology Facilities Council of the United Kingdom; the Gordon and Betty Moore Foundation; the Heising-Simons Foundation; the French Alternative Energies and Atomic Energy Commission (CEA); the National Council of Humanities, Science and Technology of Mexico (CONAHCYT); the Ministry of Science and Innovation of Spain (MICINN), and by the DESI Member Institutions: www.desi.lbl.gov/collaborating-institutions. The DESI collaboration is honored to be permitted to conduct scientific research on I’oligam Du’ag (Kitt Peak), a mountain with particular significance to the Tohono O’odham Nation. Any opinions, findings, and conclusions or recommendations expressed in this material are those of the author(s) and do not necessarily reflect the views of the U.S. National Science Foundation, the U.S. Department of Energy, or any of the listed funding agencies.
\end{acknowledgements}

%\vspace{-1.0em}
\bibliography{references}{}
\bibliographystyle{aa}

%\clearpage

\appendix

\nolinenumbers

\section{GRM using annulus method and comparison with  standard GRM Map}
\label{grm_appendix}
\begin{figure}[ht]
    \centering
    \includegraphics[width=0.95\linewidth]{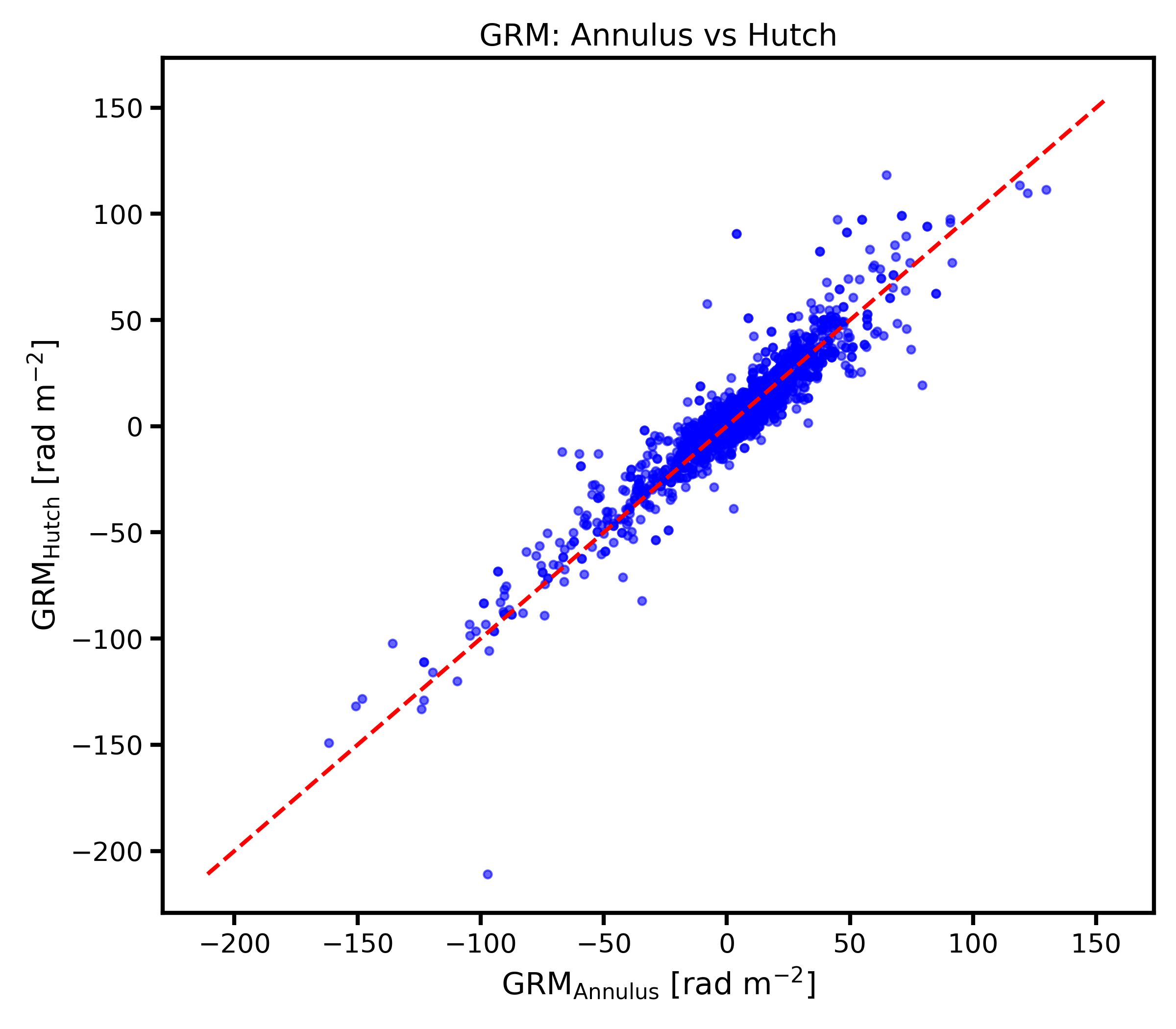}\\[0.4cm]
    \includegraphics[width=0.95\linewidth]{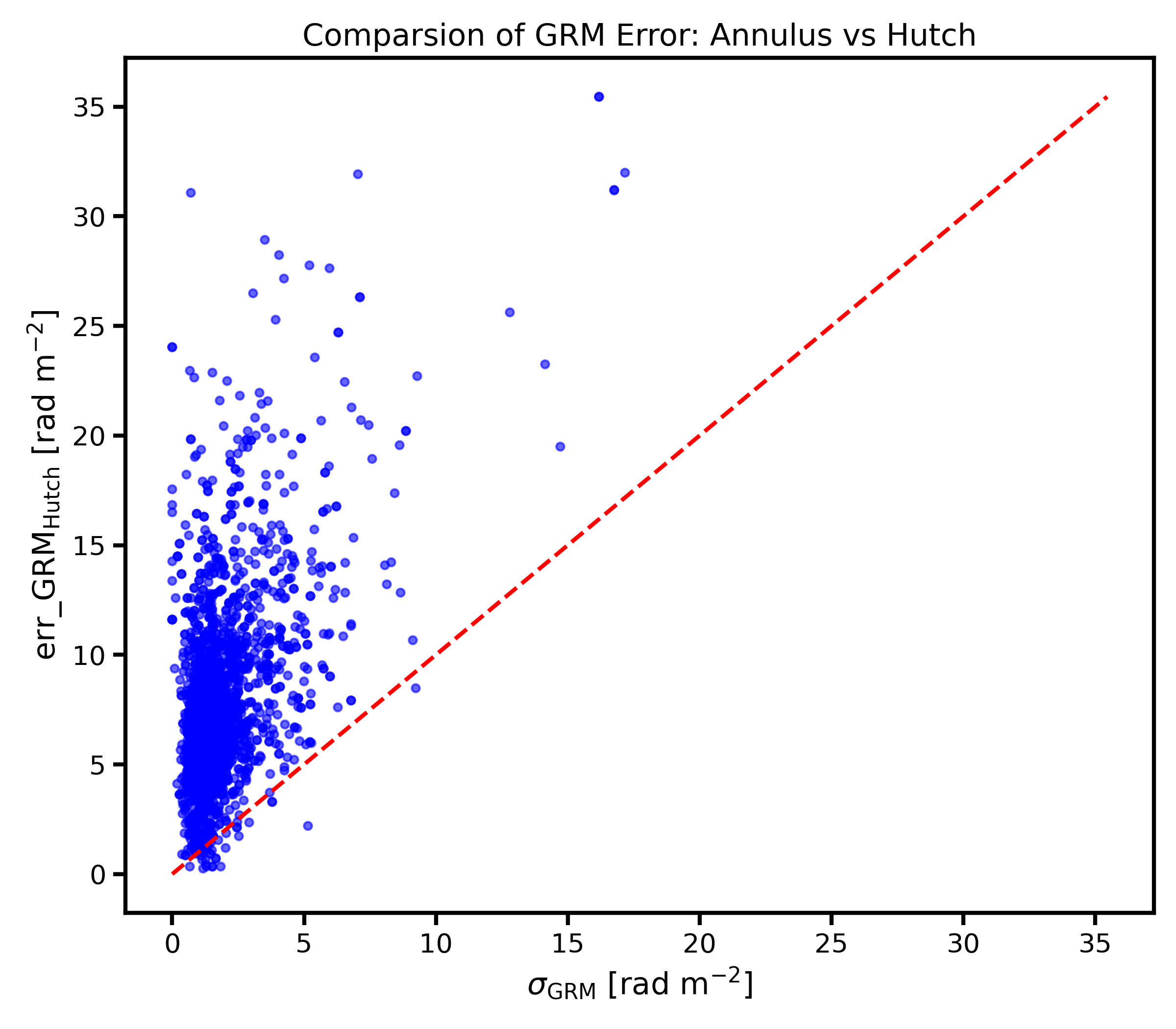}
    \caption{
    Comparison of GRM values (top panel) and their associated uncertainties (bottom panel)
    for all sightlines in our sample, derived using the \citep{Hutschenreuter_2022} map
    and our bespoke annulus-based method. 
    The red-dashed line shows the one-to-one relation. 
    }
    \label{grm_comparison}
\end{figure}

To estimate the foreground GRM for each RM component, we adopt an annulus-based method \citep{Anderson2024MNRAS}. For each target source, we construct an annulus centered on its position with an inner radius of $r_{\mathrm{inner}} = 0.2^\circ$ (at $z=1.0$, this corresponds to $\sim5.9$ Mpc using Planck cosmology~\citep{Planck2020A&A}), to avoid both self-contamination and contamination from the CGM of the foreground galaxies, and then select the 20 nearest RMs from the full sample. %reference sample. 
To minimise the influence of outliers, we discard RMs exceeding $5\sigma_{RM}$, where $\sigma_{RM}$ is the standard deviation of the selected 20 RMs, and assign their median value as the GRM at that location. The GRM uncertainty is estimated as $\sigma_{\mathrm{GRM}} = \sigma_{\mathrm{MAD}}/\sqrt{N}$, where $\sigma_{\mathrm{MAD}}$ is the median absolute deviation–based standard deviation and $N=20$ (i.e.~the statistical uncertainty).

To validate our GRM estimation procedure, we compare our locally derived Galactic RM values with the all-sky Galactic RM map of ~\cite{Hutschenreuter_2022}, which is based on a Information Field Theory reconstruction of extragalactic RM measurements. In the top panel of Fig.~\ref{grm_comparison}, we show the GRM estimated from our annular method plotted against the corresponding GRM values extracted from their map at the same sky positions. They closely follow each other, with a few outliers and a near-unity slope, demonstrating that our local annulus-based approach reliably recovers the GRM from the Milky Way.

\begin{figure}[H]
    \centering
    \includegraphics[scale=0.43]{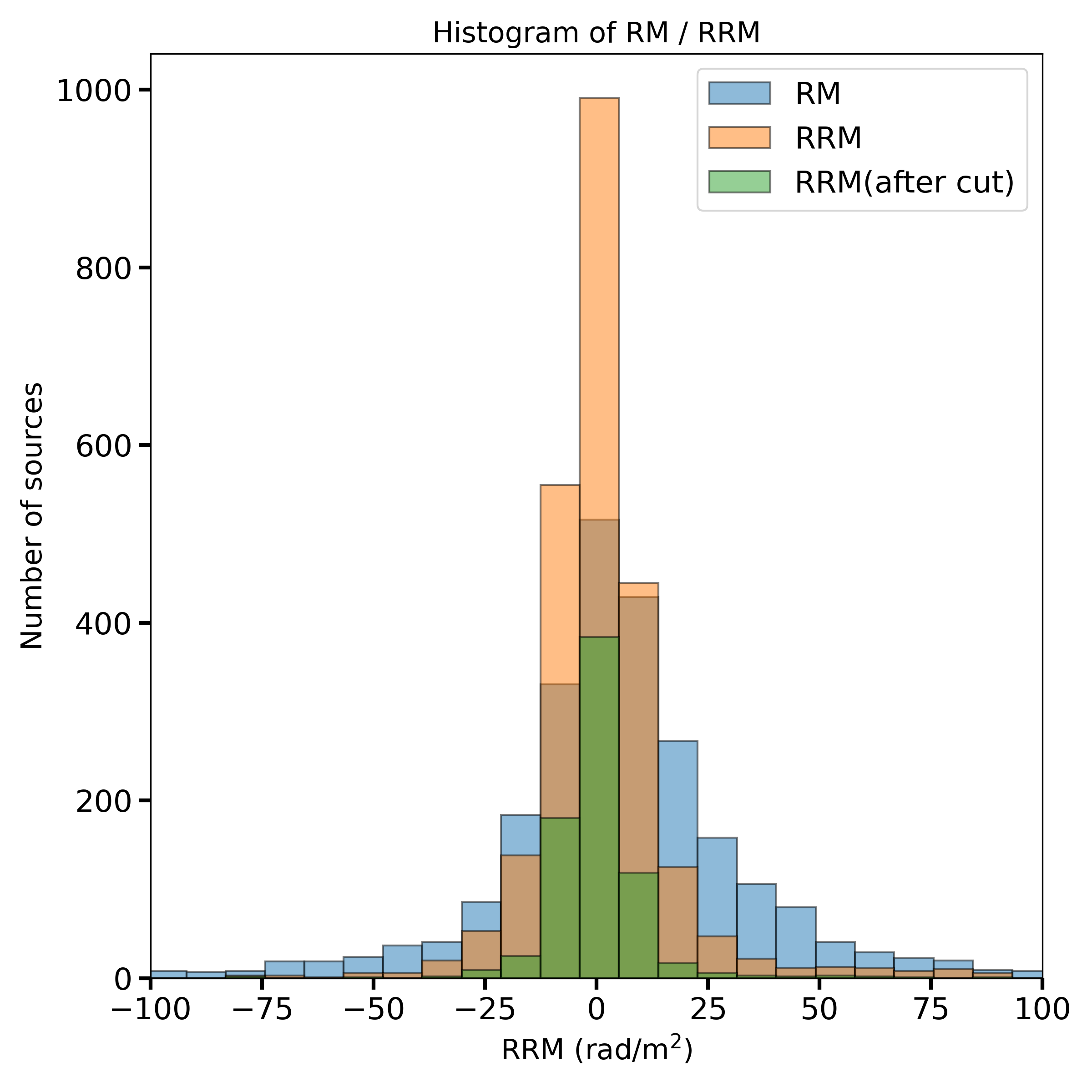}
    \caption{Histograms of RM (blue) and RRM (orange) for all sightlines in the sample show the impact of the GRM. We have also overplotted the RRM (clean sample after cuts, green) to show the impact of HI and H$\alpha$ cuts.}
    \label{histogram_rm}
\end{figure}

The bottom panel of Fig.~\ref {grm_comparison} compares the associated GRM uncertainties from our method with the uncertainty estimates provided by the ~\cite{Hutschenreuter_2022} map. We can clearly see that the uncertainty from the ~\citep{Hutschenreuter_2022} method is larger than the annulus method ($\sim$3 times on average). This behaviour is expected, as their map represents a smoothed Galactic foreground model with a lower spatial sampling of input data points than we have used. Therefore, we use the annulus method in our analysis, and the resulting RRMs are not dominated by foreground systematics. The histogram shown in Fig.~\ref{histogram_rm} also highlights the same. 

\section{Structure function Analysis}
\label{sf}
\begin{figure*}[h]
    \centering
    \includegraphics[width=0.47\linewidth]{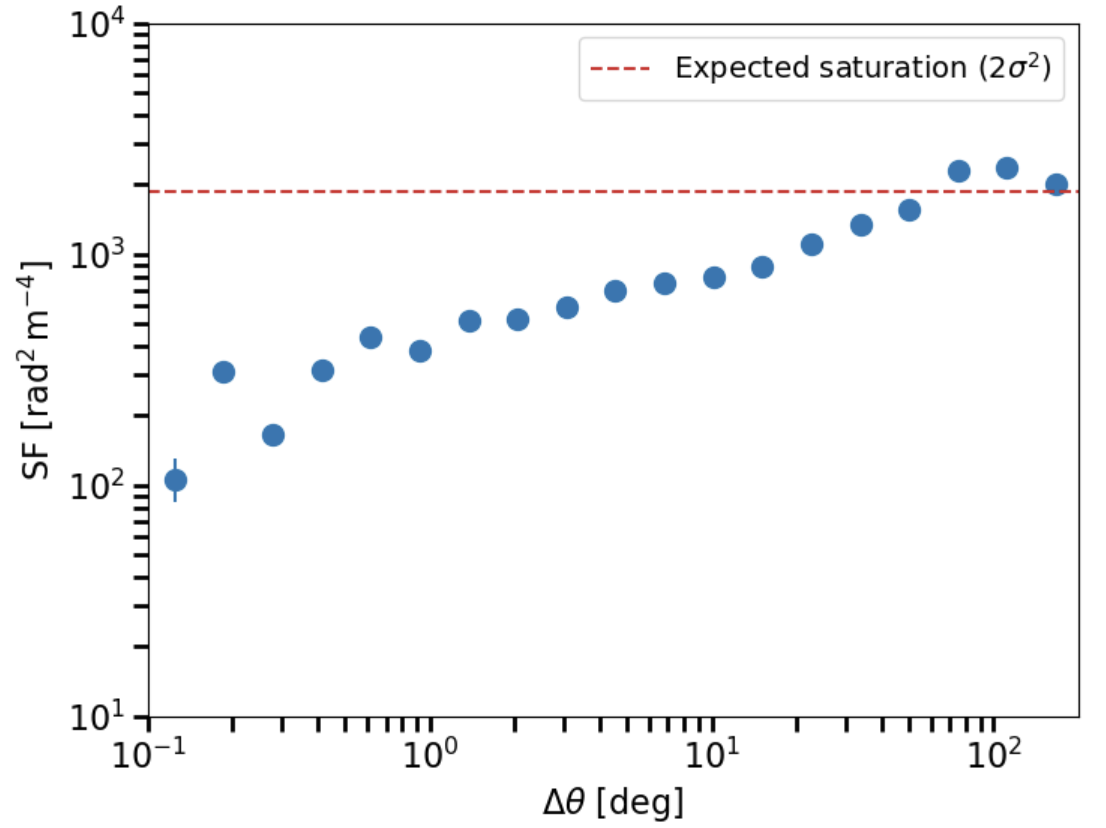}
    \includegraphics[width=0.47\linewidth]{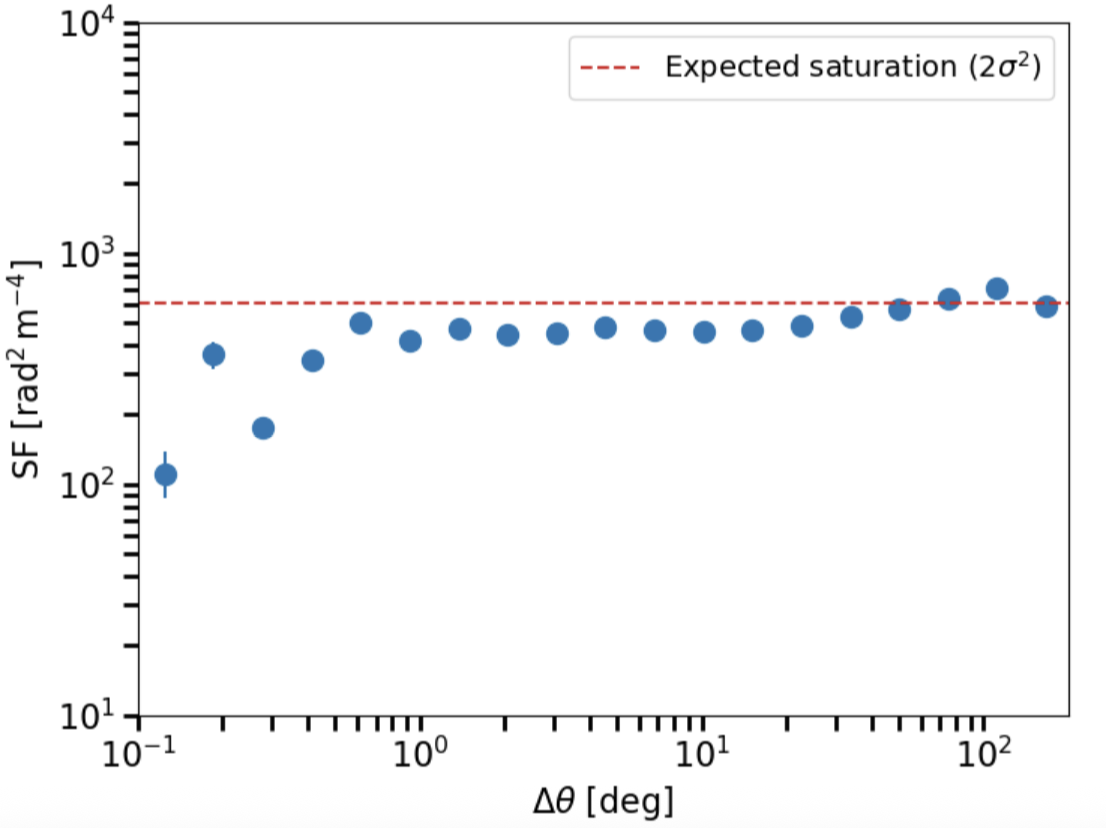}
    \caption{
    Structure function of RM as a function of angular separation (left panel), for RRM (right panel).}
    \label{sf}
\end{figure*}
To assess the impact of GRM removal on the RM, we computed the two-point structure function for both the RM and RRM using the full sample, prior to applying any H\textsc{i} or H$\alpha$ threshold criteria. The resulting structure functions are shown in Fig.~\ref{sf}. A direct comparison between the RM and RRM structure functions highlights that the majority of the scale-dependent GRM contributions are removed; however, at scales smaller than \textbf{$\sim 0.7$}\degree, their effects remain.

\section{HI and H$\alpha$ threshold} 
\label{hi_alpha_appendix}

\begin{figure*}[h]
    \centering
    \includegraphics[width=0.47\linewidth]{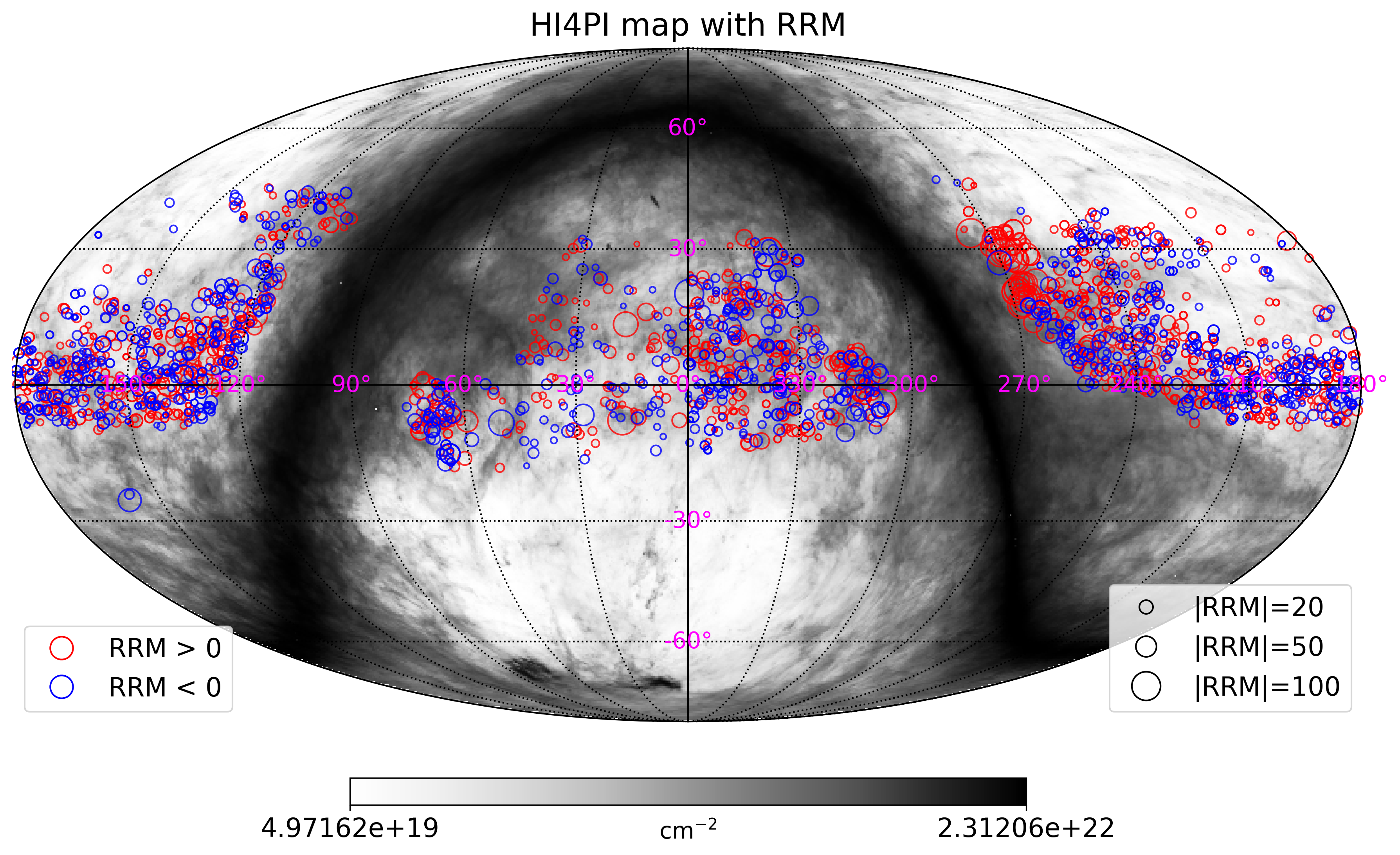}
    \includegraphics[width=0.47\linewidth]{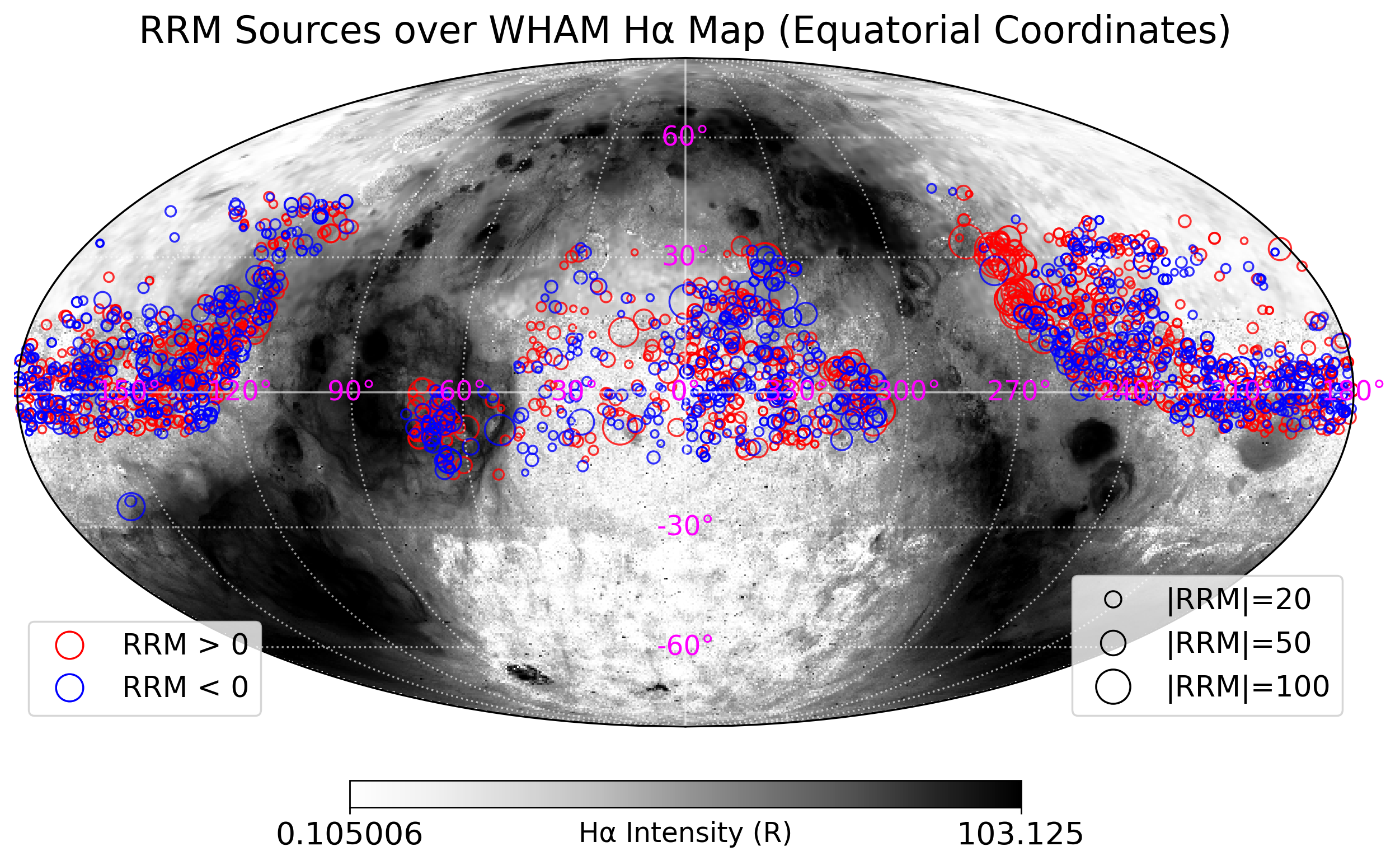}
    \caption{\emph{Left panel:} All-sky map of HI column density, overlaid with the full sample RRM ($+$ve with red circle and $-$ve with blue circle), where circle size indicates the |RRM|. \emph{Right panel:} Same as left but with H$\alpha$ intensity in the background.}
    \label{full_sample}
\end{figure*}

\begin{figure}[h]
    \centering
    \includegraphics[width=1\linewidth]{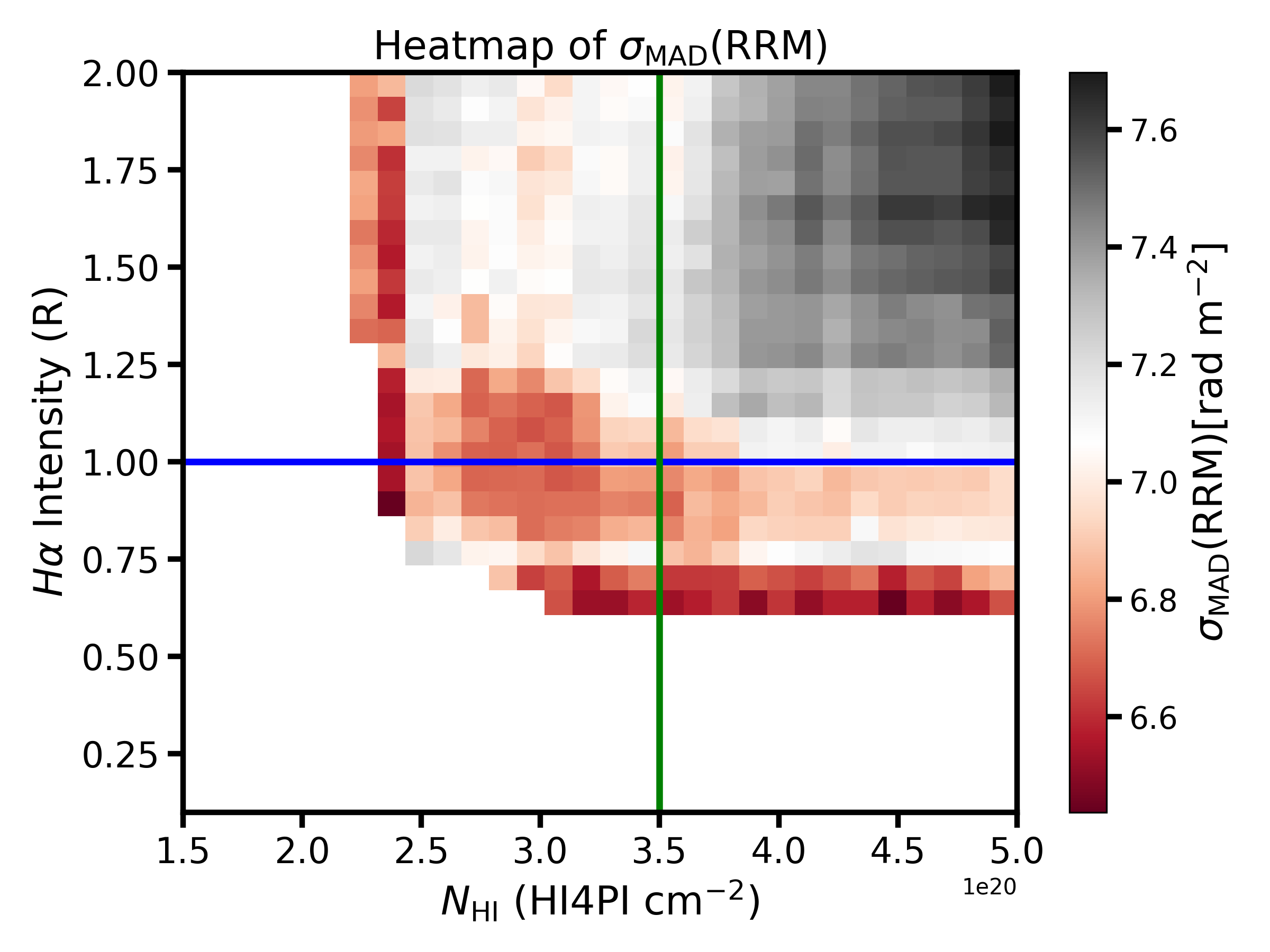}
    \caption{Map of the $\sigma_{\rm MAD}$ of RRM for various H\textsc{i} column density and H$\alpha$ intensity cutoffs for the total clean sample with and without \mgii \ absorbers. For each cutoff in both axes, a minimum of 100 \mgii \ sightlines is required, leading to missing pixels at low H$\alpha$ and H\textsc{i}. The vertical green and horizontal blue lines mark the thresholds N$_{\rm HI}$ $= 3.5 \times 10^{20}\ {\rm cm}^{-2}$ and H$\alpha$ = 1.0 R, respectively.}
    \label{sigma_mad_hi_halpha_main}
\end{figure}

In Fig.~\ref{full_sample}, we overlay the full sample of sightlines on the \textsc{HI4PI} H\textsc{i} column density map using the HI4PI map from~\cite{HI4PI2016} (in the left panel) and H$\alpha$ intensity using Wisconsin H-Alpha Mapper (WHAM\footnote{https://www.astro.wisc.edu/research/research-areas/galactic-astronomy/wham/}; \citealt{Haffner2003}) (in the right panel) to visually assess the correspondence between the neutral and ionised gas distribution and the observed RRMs. The full sample RRMs are shown as red (positive) and blue (negative) circles, with the circle size representing the absolute value of the RRM. We note that a subset of sightlines passes through high–neutral and ionised regions, which are likely to host stronger magnetic fields and can therefore have a significant impact on the measured RMs.

\begin{figure}[h]
    \centering
    \includegraphics[width=1\linewidth]{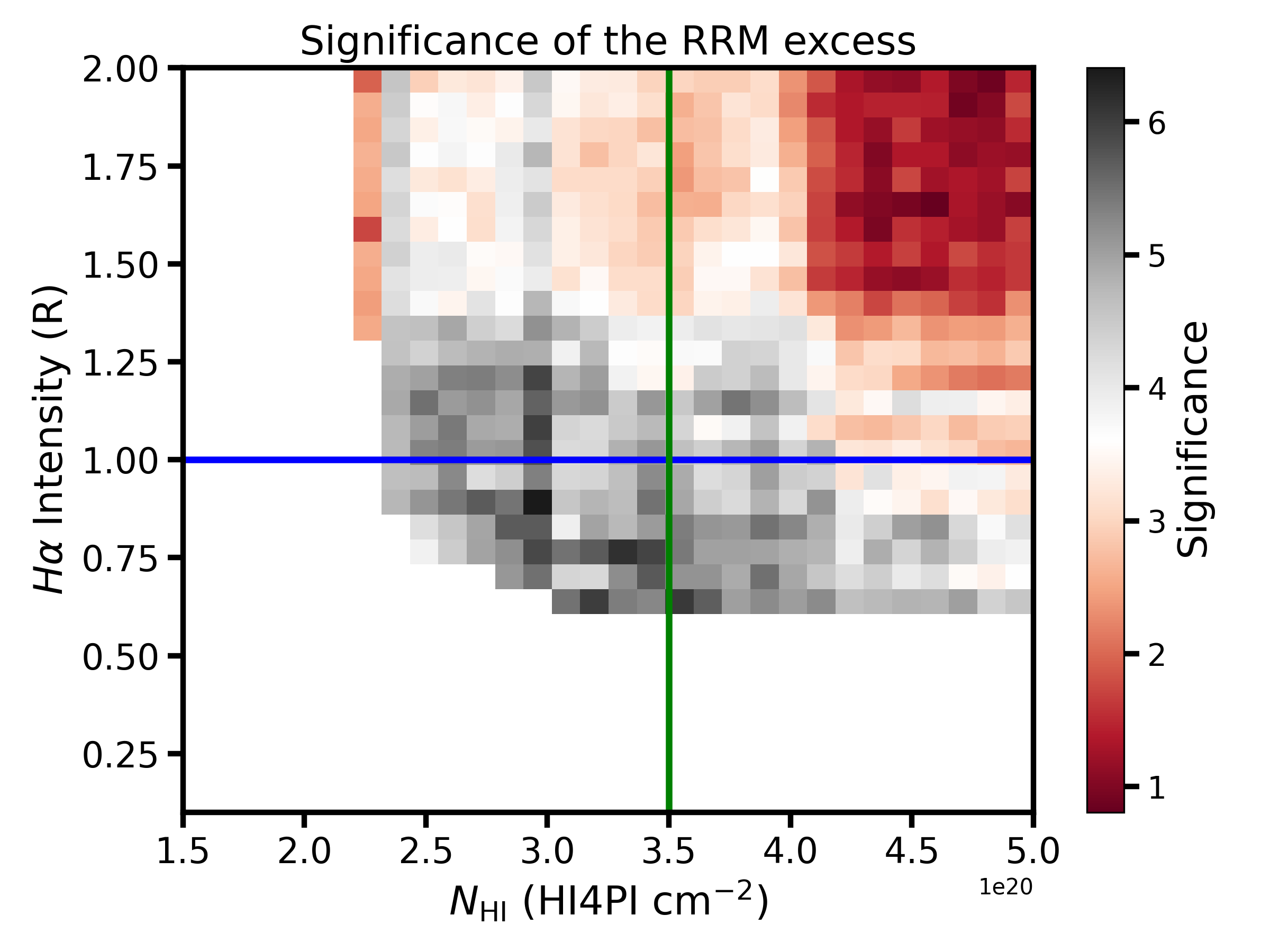}
    \caption{Map of the statistical significance of RRM excess shown in Fig.~\ref{rrm_excess_hi_halpha_main} for various H\textsc{i} column density and H$\alpha$ intensity cutoff.}
    \label{siginficane_excess_hi_halpha_main}
\end{figure}

To investigate the impact of high H\textsc{i} column density and H$\alpha$ regions on our analysis, we analysed the H\textsc{i} and H$\alpha$ intensity from the \textsc{HI4PI} and WHAM survey for each line of sight in our sample. We then computed the $\sigma_{\rm MAD}$ of RRM (combined sample of with and without \mgii~systems) as a function of H\textsc{i} column density and H$\alpha$ intensity, using sources below the different cutoffs, as shown in Fig.~\ref{sigma_mad_hi_halpha_main}. The resulting trends show a clear increase in $\sigma_{\rm MAD}$ with increasing N$_{H\textsc{i}}$ and H$\alpha$ intensity, indicating that regions with high neutral and ionised gas density exhibit significantly enhanced Faraday dispersion. This increase of $\sigma_{\rm MAD}$ (even though only 15\% from lowest to highest values) is significant enough to affect the analysis. In particular, it can affect the excess results and their uncertainty. These results demonstrate that high–H\textsc{i} and H$\alpha$ regions can substantially contaminate the extragalactic RM signal, motivating the need for an H\textsc{i} and H$\alpha$-based threshold in our analysis. We have estimated the RRM excess map as shown Section~\ref{excess_1} and its significance map is shown in Fig.~\ref{siginficane_excess_hi_halpha_main}.

\begin{figure}
    \centering
    \includegraphics[width=1\linewidth]{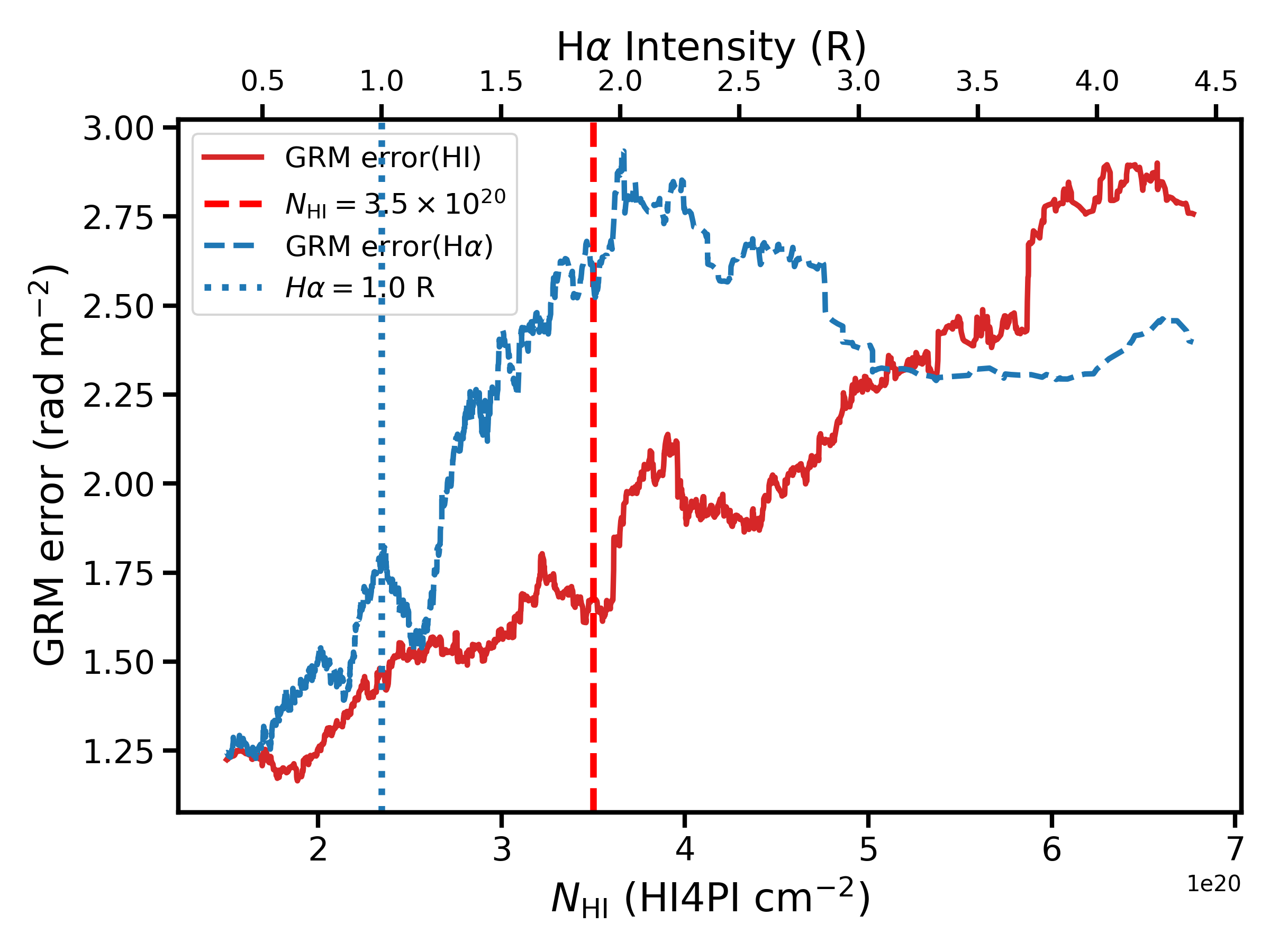}
    \caption{The GRM error plotted against  H\textsc{i} column density (bottom x-axis), and  H$\alpha$ intensity (top x-axis). Vertical dashed red and dotted blue lines mark the H\textsc{i} and H$\alpha$ thresholds, respectively.}
    \label{grm_err_hi_halpha}
\end{figure}

To determine an appropriate cutoff, we examined the GRM uncertainty as a function of H\textsc{i} column density and H$\alpha$ intensity for the full sample, as shown in Fig.~\ref{grm_err_hi_halpha}. We find that the GRM uncertainty increases systematically with increasing H\textsc{i} density and H$\alpha$ intensity. To retain sightlines with relatively low GRM uncertainty, we impose H\textsc{i} and H$\alpha$ thresholds at which the GRM uncertainty falls below the mean uncertainty of the full sample. This yields a cutoff of $N_{\rm HI} = 3.5 \times 10^{20} \mathrm{cm}^{-2}$, and H$\alpha$ = 1 R \textbf{($\mathrm{Rayleigh} = \frac{1}{4\pi} \times 10^{10} \ \mathrm{photons\ s^{-1}\ m^{-2}\ sr^{-1}}$)}, resulting in a reduced sample of 757 sightlines. This selection minimises contamination from turbulent or multiphase Galactic regions and enables a more robust interpretation of the extragalactic RRM statistics.

\begin{figure}
    \centering
    \includegraphics[width=1\linewidth]{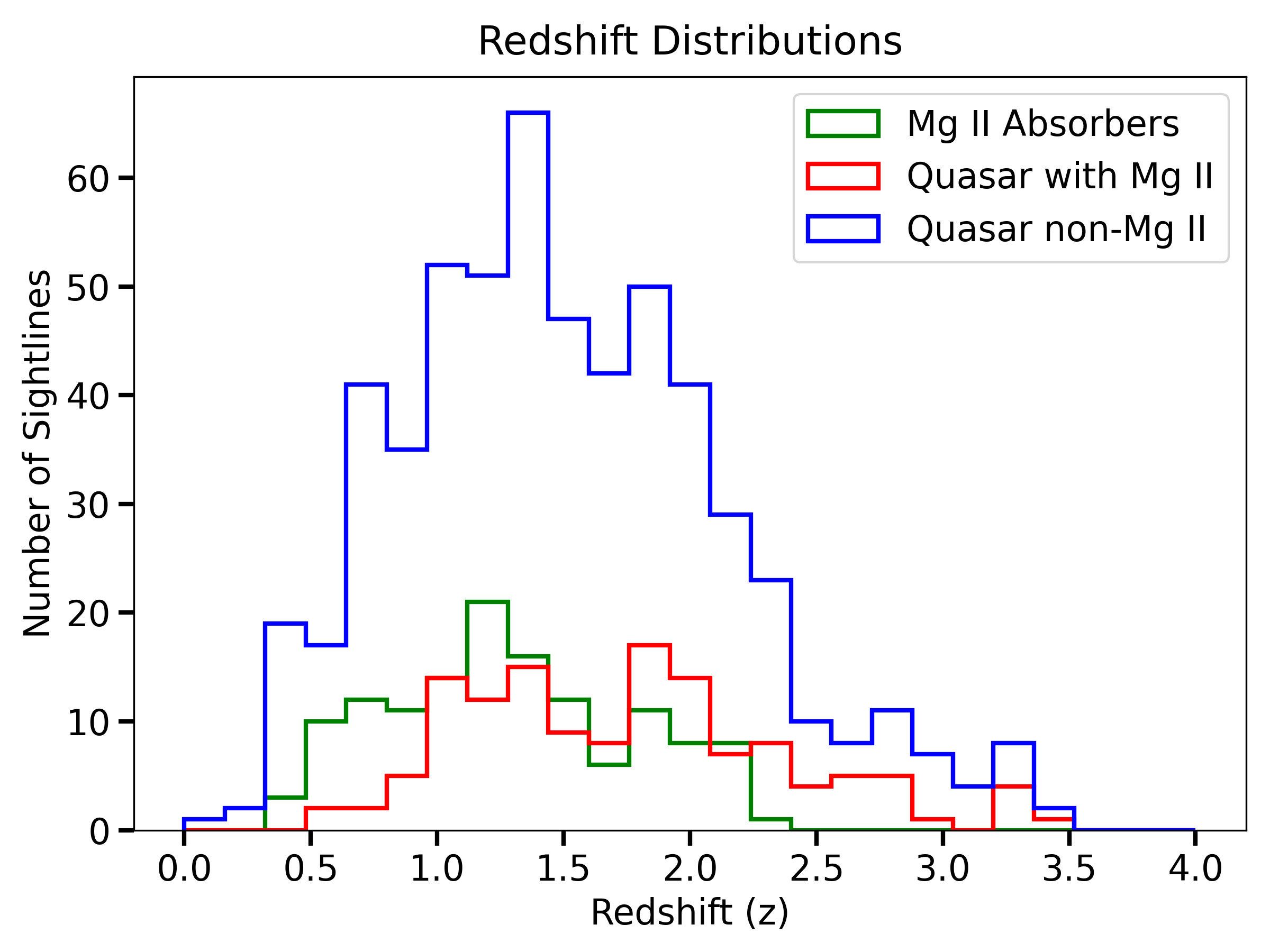}
    \caption{Histogram of the quasars’ redshift for the clean subsamples with and without \mgii \, absorbers, along with the redshift of the \mgii \, absorbers.}
    \label{z_hist}
\end{figure}

\section{HI and H$\alpha$ correlation}

\begin{figure}[h]
    \centering
    \includegraphics[width=1.0\linewidth]{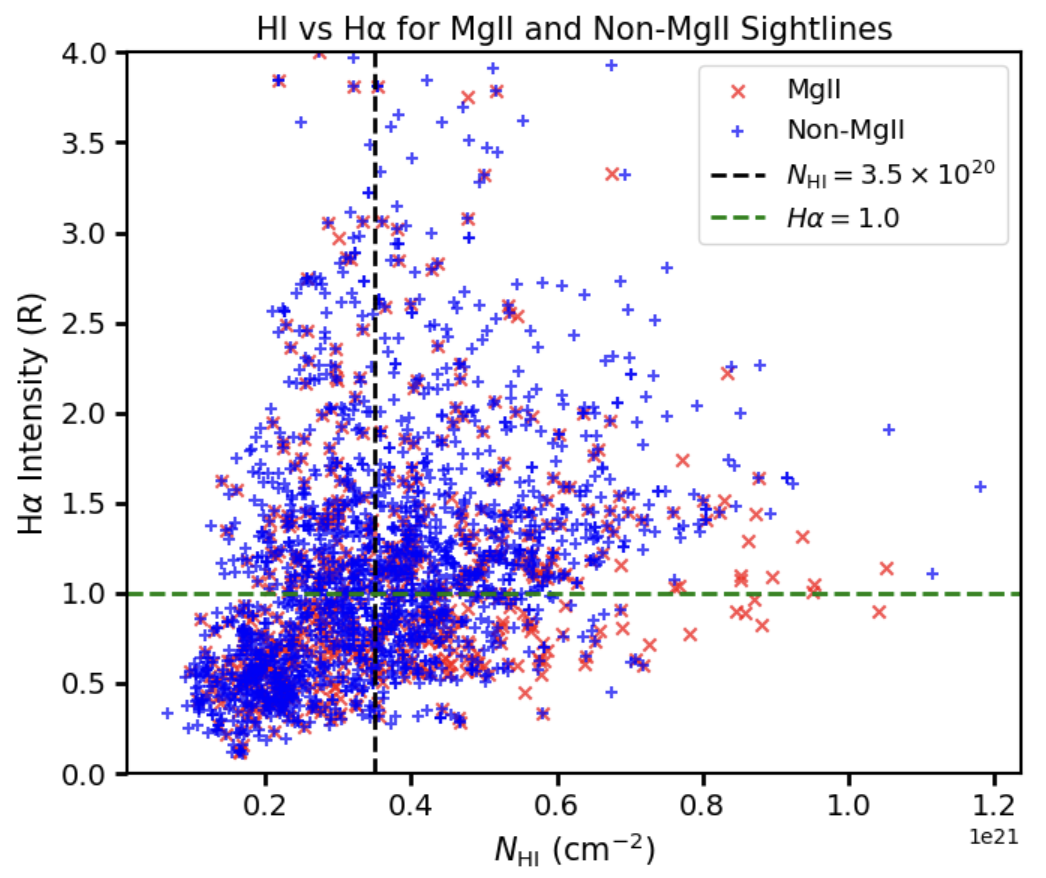}
    \caption{Distribution of H\textsc{i} column density and H$\alpha$ intensity for the full sample. The dashed vertical and horizontal lines indicate the adopted thresholds. The clean sample is represented by sources lying below both cutoffs.}
    \label{fig:HI_Halpha_corr}
\end{figure}

\begin{figure}
    \centering
    \includegraphics[width=1\linewidth]{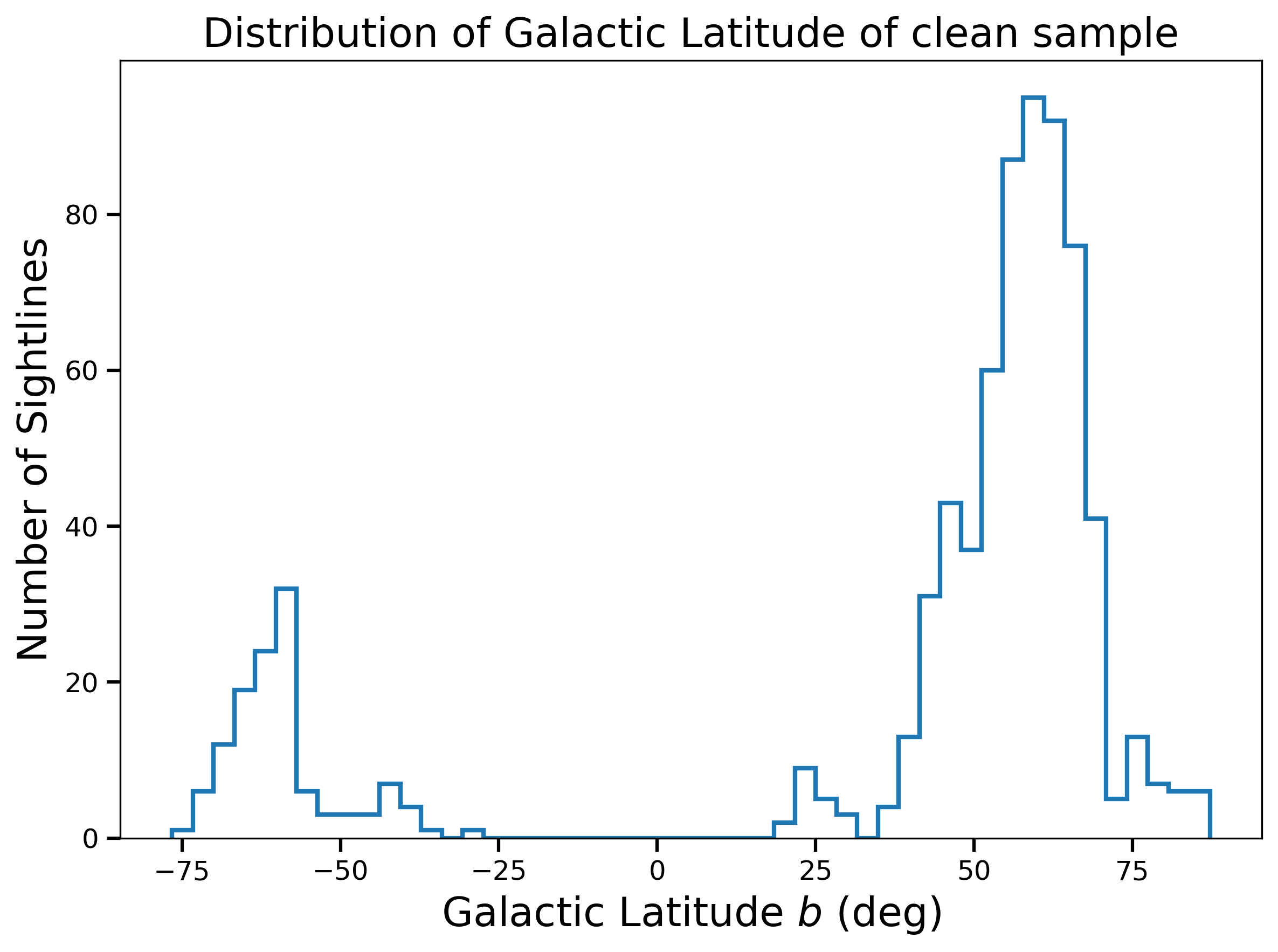}
    \caption{Histogram of Galactic latitude of the clean sample. All these sources have $|b| > 20\degree$.}
    \label{galactic_latitude}
\end{figure}

As shown in the Appendix~\ref{hi_alpha_appendix}, high-density neutral and ionised regions, quantified by H\textsc{i} column density and H$\alpha$ intensity, have a significant impact on our analysis. We also examined the one-to-one correlation between H\textsc{i} and H$\alpha$ intensity and present the corresponding scatter plot in Fig.~\ref{fig:HI_Halpha_corr}. As seen, these two quantities do exhibit a mild one-to-one correlation. Therefore, to minimise their impact on our analysis, we apply both cutoffs to filter the sightlines, as discussed in Section~\ref{clean_sample}. We have also plotted the histogram of Galactic latitude for the clean sample in Fig.~\ref{galactic_latitude}. We note that most sources lie at high Galactic latitudes, indicating minimal contamination from the Galactic plane in our analysis. 

\subsection{Classification based on Spectral Indices }
\label{spectral}
The optical and radio observations originate from different telescopes with differing angular resolutions, and therefore the observed source extents may not be perfectly aligned. The radio spectral indices, $\alpha$ ($S_\nu \propto \nu^\alpha$), provide information on the compactness of the quasars, which serves as an indicator of the alignment between the radio and optical sightlines. Accordingly, we divided the sample into three spectral categories based on the radio spectral index from~\cite{Alec2026PASA}: steep ($\alpha < -0.6$), typically associated with extended sources; flat ($\alpha \geq -0.4$), generally core-dominated; and mid-spectrum ($-0.6 < \alpha \leq -0.4$). For each category, we evaluated the $\sigma_{\rm MAD}$ of the RRM separately for sightlines with and without foreground Mg \textsc{ii} absorbers. The results are summarised in Table~\ref{tab:rrm_mgii_spectraltype}.

\begin{table}[h!]
\centering
\scriptsize
\setlength{\tabcolsep}{2pt}
\caption{RRM dispersion ($\sigma_{\rm MAD}$) and excess for Mg\,\textsc{ii} subsamples by spectral type.}
\label{tab:rrm_mgii_spectraltype}
\begin{tabular}{l l c c c}
\hline
\hline
\multirow{2}{*}{Mg\,\textsc{ii}} & \multirow{2}{*}{Spec. type} & $\sigma_{\rm MAD}$ & Sightlines & Excess \\
 &  & [rad\,m$^{-2}$] &  & [rad\,m$^{-2}$] \\
\hline
$N=0$ & Steep & $6.81 \pm 0.37$ & 282 & -- \\
      & Flat  & $6.70 \pm 0.43$ & 171 & -- \\
      & Mid   & $4.76 \pm 0.51$ & 113 & -- \\
\hline
$N>0$ & Steep & $8.10 \pm 0.65$ & 99  & $4.38 \pm 1.34$ \\
      & Flat  & $7.31 \pm 0.77$ & 57  & $2.90 \pm 2.17$ \\
      & Mid   & $7.20 \pm 1.10$ & 27  & $5.41 \pm 1.53$ \\
\hline
%$N=1$ & Steep & $8.73 \pm 0.62$ & 74  & $5.46 \pm 1.34$ \\
%      & Flat  & $7.37 \pm 0.95$ & 32  & $3.05 \pm 2.47$ \\
%      & Mid   & $6.63 \pm 1.13$ & 27  & $4.64 \pm 1.70$ \\
%\hline
%$N>2$ & Steep & $11.32 \pm 2.03$ & 11 & $9.04 \pm 2.56$ \\
%      & Flat  & $9.42 \pm 3.13$  & 5  & $6.62 \pm 4.48$ \\
%      & Mid   & $7.50 \pm 2.54$  & 5  & $5.79 \pm 3.32$ \\
%\hline
\end{tabular}
\label{tab:rrm_mgii_spectraltype}
\end{table}

We find that the RRM excess in the subsample with Mg \textsc{ii} absorbers is present in both the steep- and flat-spectrum sources, which is contrary to the earlier findings of \citet{malik_role_2020}. In that work, the RRM excess associated with foreground Mg \textsc{ii} absorbers was reported only for core-dominated sources. However, due to the relatively low resolution of the radio and optical data used in that study, the positional association was limited to a separation of $\sim 7^{\prime\prime}$, which may have affected the morphological classification. With the advent of high-resolution optical data from DESI and radio data from ASKAP, the positional separation between radio and optical counterparts is reduced to $\sim 3^{\prime\prime}$, allowing more reliable associations. As a result, we now find that sightlines classified as steep-spectrum sources, which were previously considered to arise from extended emission regions, may also lie close to optical counterparts and thus pass through foreground galaxies. This further strengthens the evidence for magnetised plasma in the halos of intervening galaxies.

\end{document}